\documentclass[12pt,preprint]{aastex}
\usepackage{graphicx,natbib}
\usepackage{emulateapj5}
                                                                                 
\begin{document} 
                               
\title{A panoramic view of the Milky Way analogue NGC~891}
                                            
\author{M. Mouhcine\altaffilmark{1},R. Ibata\altaffilmark{2}, M. Rejkuba\altaffilmark{3}}
\altaffiltext{1}{Astrophysics Research Institute, Liverpool John Moores University, 
Twelve Quays House, Egerton Wharf, Birkenhead, CH41 1LD, UK. }
\altaffiltext{2}{Observatoire Astronomique de Strasbourg (UMR 7550), 11, rue de l'Universit\'e, 
67000 Strasbourg, France.}
\altaffiltext{3}{European Southern Observatory, Karl-Schwarzschild-Strasse 2, 
D-85748 Garching, Germany.}                                       

\begin{abstract}

Recent panoramic observations of the dominant spiral galaxies of the Local Group have revolutionized 
our view of how these galaxies assemble their mass. However, it remains completely unclear whether 
the properties of the outer regions of the Local Group large spirals are typical. Here, we present  the 
first panoramic view of a spiral galaxy beyond the Local Group, based on the largest, contiguous, 
ground-based imaging survey to date resolving the stellar halo of the nearest prime analogue of 
the Milky Way, NGC~891 ($D\approx 10$~Mpc). The low surface brightness outskirts of this galaxy 
are populated by multiple, coherent, and vast substructures over the ${\rm \sim 90\,kpc \times 90\,kpc}$ 
extent of the survey. These include a giant stream, the first to be resolved into stars beyond the 
Local Group using ground-based facilities, that loops around the parent galaxy up to distances of 
$\sim 50\,$kpc. The bulge and the disk of the galaxy are found to be surrounded by a previously 
undetected large, flat and thick cocoon-like stellar structure at vertical and radial distances 
of up to $\sim 15\,$kpc and $\sim 40\,$kpc respectively. 
\end{abstract}
\keywords{galaxies: halos -- galaxies: stellar content -- galaxies: individual (NGC~891) }


\section{Introduction}

In recent years, it has been increasingly recognized that many of the clues to the problem of 
galaxy formation are preserved in galaxy outskirts \citep[e.g.][and references therein]{johnston08}. 
The current consensus is that large spirals begin as small fluctuations in the early Universe and 
grow by in situ star formation and hierarchical merging \citep[e.g.][]{wr78}. Subsequently, once 
a spiral is the dominant component in such mergers, it continues accreting and disrupting 
sub-halos falling into its potential well. With the accumulation of accretion and disruption events, 
massive spirals build up stellar and dark matter halos \citep{abadi06}. The complete disruption 
of the sub-halos may take several orbits, distributing thus the tidally stripped stars over a broad 
range of distances from the main galaxy. As galaxies are predicted to assemble their outskirts 
from the disruption of a large number of satellites, these regions are expected to possess 
significant density and chemical substructures \citep{font06}. 

Observationally however, the nature and the origin of those regions remains elusive. 
Much of what we know about their properties is based on observations of the Local Group 
massive spirals \citep[e.g.][and references therein]{fb02}. Recent surveys find evidence that 
the Galaxy stellar halo is divisible into two components, with a moderately flat inner regions 
showing a modest prograde rotation, whereas the outer regions are less chemically evolved 
and exhibit a nearly spherical distribution with a retrograde rotation \citep{carollo08}. 
Wide-field imaging data indicates that the stellar halo of the Galaxy is highly structured 
\citep{ibata03,yanny03,belokurov07}, suggesting that a large fraction of the Halo has 
been accreted from satellites \citep{bell08}. The outer regions of Andromeda have been recently 
observed to contain even more substructure and streams than observed around the Milky Way 
\citep{ibata07}, suggesting that its accretion history may have been more active than the 
suspected quiet one of the Milky Way \citep{mouhcine05a,hammer07}. It is completely unknown 
however if these properties are generic features of the outskirts of spirals, or are reflecting 
peculiar assembly histories.

We have seen recently a dramatic progress in large-scale mapping of the Milky Way and 
Andromeda, with a number of large surveys measuring photometric, kinematic, and chemical 
properties of individual stars over a wide galactic volume. Although those  surveys will 
significantly advance our understanding of the assembly of these galaxies, it cannot be 
assumed that a sample of two galaxies will provide the definitive solution to the nature and 
origin of the stellar content in the outskirts of spirals. The fundamental next step to fully exploit 
the Local Group surveys and thereby to establish a comprehensive picture of the assembly 
histories of spirals is to determine whether the Local Group massive spirals are suitably typical 
by studying giant spirals beyond the Local Group. 

A number of surveys of the low surface brightness outskirts of spirals beyond the Local Group 
have been conducted recently. These surveys were however either sampling limited galactic 
volumes \citep{mouhcine05b,blandhawthorn05,mouhcine07,dejong07}, or too shallow to detect 
the old stellar tracers \citep{davidge06,davidge07}, thus severely hampering their impact. 
Measurements of galaxy outskirts have been also attempted using integrated light 
\citep{morrison94}, and have succeeded on detecting low surface brightness tidal streams in 
the outskirts of a few nearby disk galaxies \citep{martinez-delgado08,martinez-delgado09}. 
Those measurements are however affected by large uncertainties 
\citep{dejong09,martinez-delgado08}, and are able to detect stellar structures down to a surface 
brightness limit of ${\rm \mu_{I} \sim 28\, mag\, arcseec^{-2}}$ at the faintest \citep{zheng99}, 
many magnitudes brighter than the bulk of substructure detected in the outskirts of the Local 
Group spirals \citep{belokurov07,ibata07}.

Characterizing comprehensively the outskirts of spirals beyond the Local Group requires resolving 
panoramically the stellar content of those regions, giving access to the extremely low surface 
brightness structures. The required observations are however extremely challenging due to the 
photometric depth one has to reach, i.e., $I\sim 26.-28.5$, to resolve old giant stars in galaxies 
beyond the Local Group, restricting this approach (using the present-day instrumentation) to the  
small number of spirals closer than $\sim 10-12\,$Mpc. 
 
Because the theoretical predictions are inherently statistical in nature, and due to the stochastic 
nature of halo formation, constraining the assembly history of galaxy outskirts must rely in large 
part on measuring the demographics of their stellar populations. To this end, we have initiated 
a survey to resolve panoramically the stellar content of galaxy outskirts for a sample of nearby, 
i.e., up to $\sim10$ Mpc, highly inclined spirals, distributed over a broad range of masses, 
i.e., circular velocities ranging from ${\rm \sim 80\,km\,s^{-1}}$ up to ${\rm \sim 230\,km\,s^{-1}}$, 
and morphologies, i.e., ranging from Sa to Sd. 

As part of this effort, we have targeted the edge-on galaxy NGC~891, often considered as 
the nearest prime analog of the Galaxy. Our group recently used deep optical imaging data, 
obtained with the Advanced Camera for Surveys on board the Hubble Space Telescope 
(HST/ACS), to investigate the properties of the resolved extra-planar stellar populations over 
the south-east quadrant of NGC~891, extending up to $\sim10$\,kpc from the galactic plane
 \citep{mouhcine07,ibata09,rejkuba09,harris09}. 
Succinctly summarized, those studies indicate that the thick disk of NGC~891 shares comparable 
structural and chemical properties to its Galactic counterpart. The stellar populations beyond 
the thick disk appear to possess significant small-scale variations in the stellar metallicity. 
Interestingly, those regions are found to be dominated by stars significantly more chemically 
enriched than those populating the regions of comparable heights from the plane of the Galaxy.

In the present contribution, we report the first results of our survey. Detailed analysis of the 
properties of the stellar content of different galactic components, the search for substructures, 
the determination of metallicity distribution functions and their spatial variation, and the 
properties of the globular cluster system will be reported in forthcoming papers. 
The layout of this paper is as follow; in \S~\ref{data} we present briefly the data set. 
In \S~\ref{analysis} we report the discovery of a new morphological structures around NGC~891, 
and discuss the implications of our finding for the formation and evolution of spiral galaxies. 

Throughout the paper we use an intrinsic distance modulus for NGC~891 of 
$(m-M)_{\circ} = 29.94$, with the tip of the red giant branch located at  $I\,=\,25.84 \pm 0.04$ mag
\citep{mouhcine07,tg05}. NGC~891 is located at relatively low Galactic latitude 
($\ell=140.38^\circ$, $b=-17.42^\circ$), and therefore it suffers from significant (though not large) 
extinction from foreground dust: $E(B-V)= 0.065$ \citep{schlegel98}.

\section{Data}
\label{data}

The Subaru Prime Focus Camera (Suprime-Cam) on the 8.2-m Subaru Telescope is a  mosaic 
camera of ten-chip $2048 \times 4096$ charge-coupled devices, which 
covers a ${\rm 34\,arcmin \times 27\,arcmin}$ field of view with a pixel scale of 
0.20\,arcsec \citep{miyazaki02}. The instrument was used to image the outer regions of NGC~891.
In the following, we will describe briefly the observations and the data reduction, the full details 
will be reported elsewhere.

The observations were taken in the Johnson visual V-band and Gunn $i$-band. We have obtained 
a total of 10 hours of good quality data in the V-band and 11.27 hours in the $i$-band, with seeing 
better than 0.6 arcsec, allowing us to resolve approximately the brightest two magnitudes of the 
red giant branch (RGB hereafter) of metal-poor stellar populations at the distance of NGC~891. 
By covering a vast ${\rm \sim 90\,kpc \times 90\,kpc}$ region around NGC~891, the survey allows 
us to distinguish local density enhancements from large-scale structure of the halo of the galaxy 
and/or the foreground distribution of stars.
 
The images were pre-processed using the the Cambridge Astronomical Survey Unit photometric 
pipeline. The final stacks were created by summing all pixels, weighted by the estimated seeing 
on each frame. The DAOPHOT/ALLSTAR software suite \citep{stetson87} was used to detect all 
sources down to $3\sigma$ above the sky. Isolated bright stars were identified over the frame to 
serve as point spread function (PSF) templates. The PSF was modelled as a Moffat function; 
experiments showed that allowing the PSF to vary spatially over the field did not improve 
significantly the fit to bright stars, so we adopted a spatially constant PSF.  Finally, the instrumental 
magnitudes were shifted to agree with the calibrated photometry presented in \citet{rejkuba09}.

The analysis of the photometric errors indicates that the average $i$-band error varies from 
$\sigma_i  \approx 0.05$ at $i \approx 25.8-26.0$, i.e., the bright end of the RGB, to 
$\sigma_i  \approx 0.15$ at $i \approx 26.8-27.0$, i.e., approximately a magnitude below 
the tip of the RGB, while the average color error varies from $\sigma_{(V-i)}  \approx 0.12$ 
to $\sigma_{(V-i)}  \approx 0.18$ within the same magnitude range. To estimate the photometric 
completeness of the survey, we used the artificial star simulations performed for our analysis of 
the HST/ACS deep images of NGC~891, and described in full detail in 
\citet{rejkuba09}. By comparing directly the number of stars detected in both HST/ACS and 
Supreme-Cam images as a function of magnitude and spatial density of RGB stars, it is 
possible to compute the photometric completeness of our survey. For stars located above 
${\rm \sim 6\,kpc}$ from the galactic plane, i.e., regions where the crowding is not at all 
important (see \citet{rejkuba09} for more details), the data are well above $\sim 80$ per cent 
complete for the upper one magnitude of the giant branch, i.e., $i_{\circ} \la 27$.

Objects were classified as artefacts, galaxies or stars according to their morphological structure 
on the images. The point source catalog consists of NGC~891 RGB stars, NGC~891 asymptotic 
giant branch (AGB hereafter) stars, Galactic foreground stars, and unresolved background 
galaxies. In order to isolate stars to probe the spatial distribution of the stellar populations in 
the outskirts of NGC~891, we applied a series of magnitude and color cuts. RGB stars are 
defined as those with foreground extinction-corrected $i$-band magnitude fainter than 25.8 and 
brighter than 27, and with (V-$i$)$_{\circ}$ color redder than 1.1, while AGB stars are selected 
as those with foreground extinction-corrected $i$-band magnitude ranging between 25.0 and 
25.8 and foreground extinction-corrected (V-$i$)$_{\circ}$ colors redder than 1.2.

\section{The panoramic landscape around NGC~891}
\label{analysis}

Fig.~1 shows the surface density map of RGB stars across the surveyed area. 
Numerous enhancements of the surface density of RGB stars are visible with the most striking 
being the large-scale complex of arcing loops and streams wrapped around the galaxy. 
About a half-dozen arc-like features are visible in the density map of old stars, extending up 
to $\sim 40\,$kpc west and $\sim 30\,$kpc east of the galactic stellar disk. On the left side, 
corresponding to north-eastern part of the galaxy, it is possible to trace the full extension of 
a giant loop turning around, falling toward the plane of the galaxy, and extending further to 
the south. The spatial density of AGB stars does not exhibit however any obvious 
enhancements associated with the large-scale complex of streams of old stars. 
Note that there is no evidence for a significant clumping in the distribution in of extended 
sources across our field, Galaxy clusters and other large-scale structure are unlikely to 
dominate the counts at the angular scales of the observed features \citep{couch93}.

It is tempting to suggest that the large-scale network of old star streams is connected, originating 
from a single accretion event, however it is impossible to argue this firmly based on the stellar 
density map alone. Nevertheless, the arc-like features exhibit similar shapes, angular lengths, 
radii of curvature, separation from the parent galaxy, and levels of stellar density enhancements. 
In addition, the stellar loops appear to be distributed and to cross each other in a pattern 
strikingly similar to the classical rosette-shaped loops predicted for tidally disrupting dwarf 
galaxies in $N$-body simulations \citep[e.g.][]{ibata98,law05}. These properties provide 
circumstantial evidence that the arclike features around the galaxy may all come from a single 
accretion event (detailed modeling of the stellar stream will be presented elsewhere). 
The discovery of the NGC~891 giant stream in the first deep, panoramic survey of the Milky 
Way's nearest analog, together with the previous discoveries of tidal streams in the Milky Way, 
Andromeda, NGC~5907, and NGC~4013, suggests that halo substructure in the form of tidal 
streams may be a generic property of massive spirals, and that the formation of galaxies 
continues at a moderate rate up to the present day. 

Fig.~1 shows that the spatial density of old stars varies along the streams, however nowhere 
among these features is there a region dense enough that it could be identified as the 
remaining core of the disrupted progenitor dwarf satellite. The stellar streams are most likely 
the fossil remnants of a totally cannibalized system, although the core of the disrupted satellite 
could possibly be hidden behind the disk of the galaxy. Deep radio observations of NGC~891 
have reported the presence of a large gaseous filament and a number of counter-rotating 
clouds in its halo, extending to comparable distances from the disk as the stellar stream 
\citep{oosterloo07}. The structure and the kinematics of these gaseous structures appear to 
favor a scenario in which they are the results of a flyby interaction with the gas-rich satellite 
UGC~1807 \citep{oosterloo07,mapelli08}. Fig.~2 shows the color-magnitude diagram of 
stars along the streams, with fiducials of RGB sequences of three Galactic globular clusters 
from \citet{dacosta90} superimposed. The morphology of the the color-magnitude diagram 
indicates clearly that the stream progenitor contained a broad range of stellar populations, 
with a mean metallicity comparable to that of 47 Tuc, i.e., ${\rm [Fe/H]\sim-0.7}$, and ages 
older than a few Gyr with hardly any young stellar components, in sharp contrast with the 
typical stellar population of gas-rich dwarf galaxies \citep{skillman}. This indicates that it is 
highly unlikely that the large-scale complex of loops and streams in the halo of NGC~891 
could be associated with the gas-rich dwarf companion.


The second striking feature visible in the surface density map of RGB stars is the extended 
super-thick envelope surrounding the galaxy. The defining high surface brightness components 
of a spiral, i.e., the bulge and the disk, are embedded in a vast flattened cocoon-like structure 
extending vertically up to $\sim 15\,$kpc, and radially up to $\sim 40\,$kpc, with a structure 
quite unlike any classical notions of thick disk or halo. This structure is strikingly flatter than 
the inner component of the Galactic halo that dominates out to $\sim10$\,kpc above the 
disk \citep{preston91,carollo08}. An additional highly flattened component of the Galactic halo 
has been recently detected \citep{morrison09}. This halo flat component is however suspected 
to be confined within very close distances from the Galactic plane \citep{preston91,morrison09}

A stellar structure resembling the Galaxy thick disk has been resolved in NGC~891, with a 
vertical scale height of $h_Z\sim 1.45$ kpc \citep{ibata09}. Stars in this morphological structure 
show no metallicity gradients both vertically and radially, and cover a wide range in 
metallicity, similar to what is observed for the Galaxy \citep{gilmore95,ivezic08}. However, at 
vertical distances larger than $\sim 5$ kpc from the plane of the galaxy, stars belong to a more 
extended structure \citep{ibata09}, and possess a negative metallicity gradient, albeit a mild 
one \citep{rejkuba09}. The envelope surrounding NGC~891 is therefore highly unlikely an 
extension of the thick disk. Thus it appears that the flattened super-thick envelope surrounding 
the galaxy is a previously unknown component of NGC~891.

Our previous investigations of the properties of the stellar content of the south-east quadrant, 
sampling stars of the super-thick envelope, find strong evidence for large amounts of 
chemically-distinct sub-structures \citep{ibata09}. This suggests that those regions are 
populated by numerous accretion remnants that are spread over a large volume and are 
still far from being fully phase mixed. Since it is natural to expect that the stellar populations 
in this quadrant are representative of those populating the super-thick structure, this indicates 
that the stellar structure surrounding the galaxy was assembled from the tidal disruption of 
many accreted satellites.

The super-thick structure presented here is not just a peculiarity of NGC 891: a preliminary 
analysis by our group of deep panoramic observations of the nearby edge-on galaxy 
NGC~2683 shows clearly the presence of an almost identical structure surrounding the high 
surface brightness components of that galaxy. Detailed kinematical measurements have 
revealed that a number of stellar substructures present at large radii of Andromeda co-rotate 
with the inner stellar disk, and are distributed in a gigantic flattened ``extended disk'' structure, 
extending radially from the high surface brightness inner disk out to $\sim 40\,$kpc from the 
galaxy center \citep{ibata05}, i.e., strikingly similar to the radial extent of the super-thick stellar 
structure surrounding NGC 891. Given that we do not observe Andromeda edge-on, it is 
possible that the ``extended disk'' in that galaxy has a similar vertical extent to the new 
morphological structure identified in NGC~891 and NGC~2683.

The debris resulting from the disruption of a dwarf galaxy on a prograde orbit that is coplanar 
with the host galaxy disk is expected to relax into an extended rotating disk \citep{penarrubia06}. 
The detection of the Monoceros stream in the Galaxy \citep{yanny03,ibata03}, the presence of 
an extended-disk around Andromeda \citep{ibata05}, and the stellar stream around NGC~4013
\citep{martinez-delgado09}, all suspected to be the remnants of the disruption of satellites 
moving in low inclined orbits, suggest that the accretion of satellites on low-inclination and low 
eccentricity orbits may be a relatively common process in the formation of spirals \citep{pohlen04}.
Numerical simulations in the framework of the hierarchical cold dark formation scenario predict 
that large spirals experience several mergers of dwarfs with masses ranging from 1\% to 10\% 
of the host mass \citep{gao04}, with many on low-eccentricity orbits \citep{ghigna98}. 
The present work suggests that a super-thick stellar envelope formed by numerous accretions 
may be a common feature of large spirals.

\section{Acknowledgements}
This work is based on data obtained at the Subaru Telescope, operated by the National 
Astronomical Observatory of Japan.

\begin{figure}
\includegraphics[height=3.5in]{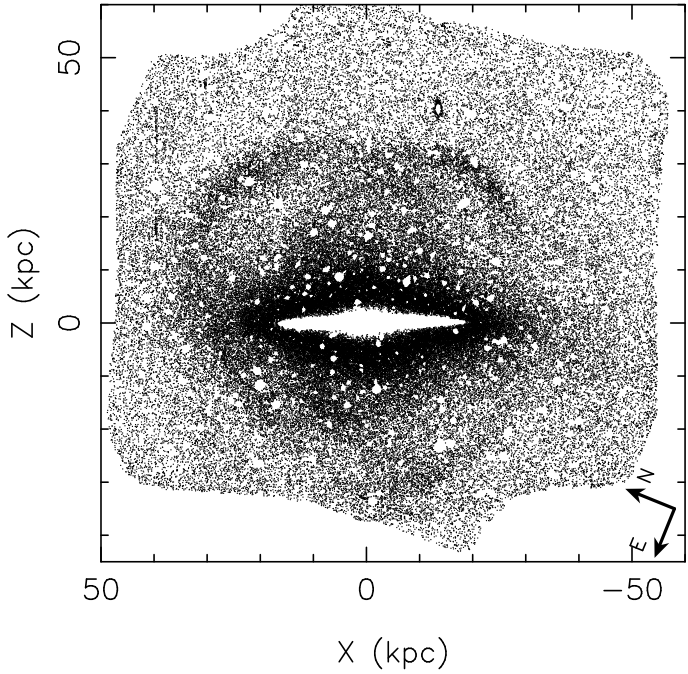}
\caption{Surface density map of RGB stars over the surveyed area around NGC~891.
The over-densities of old RGB stars detected in the present study reveal a large complex of arcing 
streams that loops around the galaxy, tracing the remnants of an ancient accretion. 
The second spectacular morphological feature is the dark cocoon-like structure enveloping the 
high surface brightness disk and bulge.}
\label{rgb_map}
\end{figure}

\begin{figure}
\includegraphics[height=3.5in]{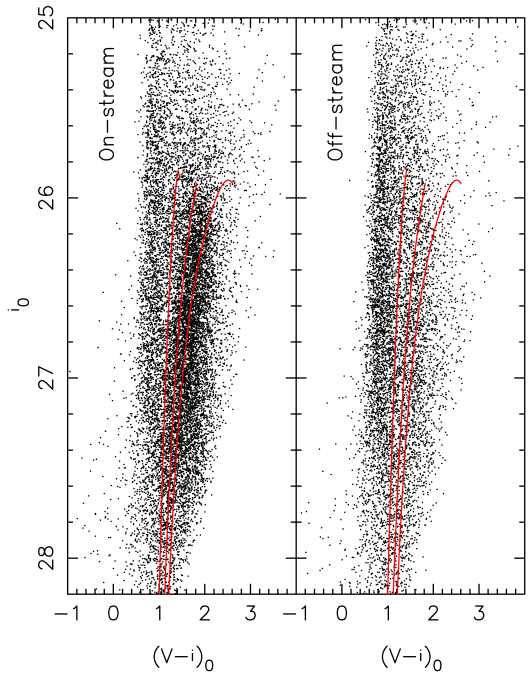}
\caption{$i_{\circ}$ vs. $(V-i)_{\circ}$ color-magnitude diagrams for a stream field (left panel) and 
an off-stream field (right panel). Superimposed on each panel are RGB tracks (shifted to the distance 
of NGC~891) of three Galactic globular clusters of different metallicities: M~15 ([Fe/H]=-2.2), 
NGC~1851 ([Fe/H]=-1.2), and 47 Tuc ([Fe/H]=-0.7) from \citet{dacosta90}.}
\label{cmd_stream}
\end{figure}

\end{document}